\documentclass{article}
\usepackage{graphicx}
\usepackage{amsmath}
\usepackage{amsfonts}
\usepackage{amssymb}
\newtheorem{theorem}{Theorem}
\newtheorem{acknowledgement}[theorem]{Acknowledgement}

\begin{document}

\title{{\Large Darboux transformation for two-level system}}
\author{V. Bagrov\thanks{On leave from Tomsk State University and Tomsk Institute of
High Current Electronics, Russia, e-mail: bagrov@phys.tsu.ru}, M.
Baldiotti\thanks{e-mail: baldiott@fma.if.usp.br}, D. Gitman\thanks{e-mail:
gitman@fma.if.usp.br}, and V. Shamshutdinova\thanks{Tomsk State University}\\ \\Instituto de F\'{\i}sica, Universidade de S\~{a}o Paulo,\\Caixa Postal 66318-CEP, 05315-970 S\~{a}o Paulo, S.P., Brazil}
\maketitle
\begin{abstract}
We develop the Darboux procedure for the case of the two-level system. In
particular, it is demonstrated that one can construct the Darboux intertwining
operator that does not violate the specific structure of the equations of the
two-level system, transforming only one real potential into another real
potential. We apply the obtained Darboux transformation to known exact
solutions of the two-level system. Thus, we find three classes of new
solutions for the two-level system and the corresponding new potentials that
allow such solutions.
\end{abstract}

\section{Introduction}

It is well-known that some complex quantum systems, with a discrete energy
spectrum, are situated in some special dynamical configuration in which only
two stationary states are important. To describe such systems one can use
appropriate models with two-level energy spectra. In a number of important
cases a model for a two-level system in a time-dependent background is based
on the following Schr\"{o}dinger equation in $0+1$ dimensions for a
two-component time dependent spinor $\Psi\left(  t\right)  $,%
\begin{align}
&  i\frac{d\Psi}{dt}=\left(  \mathbf{\sigma F}\right)  \Psi\,,\;\Psi=\left(
\begin{array}
[c]{c}%
\psi_{1}\\
\psi_{2}%
\end{array}
\right)  \,,\nonumber\\
&  \mathbf{F}=\left(  \varepsilon,0,f\left(  t\right)  \right)  \,,\;\left(
\mathbf{\sigma F}\right)  =\left(
\begin{array}
[c]{cc}%
f\left(  t\right)  & \varepsilon\\
\varepsilon & -f\left(  t\right)
\end{array}
\right)  \,. \label{1}%
\end{align}
Here $\varepsilon$ is a constant and $f\left(  t\right)  $ is a real function
(in what follows, we call $f\left(  t\right)  $ the potential)\footnote{Here
and in what follows $\hbar=c=1$, and $\mathbf{\sigma}=\left(  \sigma
_{1},\sigma_{2},\sigma_{3}\right)  $ are Pauli matrices.}. The components of
the spinor obey the set of equations%
\begin{equation}
i\dot{\psi}_{2}+f\psi_{2}=\varepsilon\psi_{1}\,,\;i\dot{\psi}_{1}-f\psi
_{1}=\varepsilon\psi_{2}\,. \label{5}%
\end{equation}
These equations and their solutions are studied in the present article.

Solutions of the two-level system with different potentials possess a wide
range of applications, e.g., in quantum optics and in the semi-classical
theory of laser. The model can be helpful to describe the behavior of molecule
beams that cross a cavity immerse in a time dependent magnetic, or electric,
field, as well as the behavior of an atom under the action of the electric
field of a laser (see, for example, \cite{Nus73}). It can be mentioned,
additionally, that two-level models are used to describe resonance absorption
and nuclear induction experiments \cite{RabRaS54}. The two-level system with
periodic (quasi-periodic) potentials $f\left(  t\right)  $ has been studied by
several authors. They have considered various approximation methods for
finding solutions of the equations (\ref{1}, \ref{5}), e.g. perturbational
expansions \cite{BarCo02}, the method of averaging \cite{BarWr00}, and the
rotating wave approximation method. For a review of these, and other methods,
see \cite{GriHa98}.

One ought to say that equations (\ref{1}, \ref{5}) can be related to
Zakharov-Shabat equations \cite{NovMaP84,Laugh93}%
\begin{equation}
i\dot{\psi}_{1}+\varphi\psi_{2}=\epsilon\psi_{1},\,\;i\dot{\psi}_{2}%
-\varphi^{\ast}\psi_{1}=-\epsilon\psi_{2}\,,\;\varphi=f_{1}+if_{2}\,.
\label{2}%
\end{equation}
Indeed, for%
\begin{equation}
f_{1}\left(  i\tau\right)  =f\left(  \tau\right)  ,\;\,f_{2}=0,\;\epsilon
=-i\varepsilon\,,\;t=i\tau,\;(\tau\mathit{\;}\mathrm{is\,\operatorname{real}%
),} \label{3}%
\end{equation}
we can write the Zakharov-Shabat equations as follows%
\begin{equation}
i\frac{d\Psi}{d\tau}=\left(  \mathbf{\sigma F}_{\mathrm{ZS}}\right)
\Psi\,,\;\mathbf{F}_{\mathrm{ZS}}=\left(  0,f\left(  \tau\right)
,\varepsilon\right)  \,,\;\Psi=\left(
\begin{array}
[c]{c}%
\psi_{1}\left(  \tau\right) \\
\psi_{2}\left(  \tau\right)
\end{array}
\right)  \,. \label{4}%
\end{equation}
The latter equation can be transformed into the equation (\ref{1}) by the
unitary transformation,%
\[
\Psi\rightarrow U\Psi\,,\;U=\frac{1}{2}\left[  1+i\left(  \mathbf{\sigma
e}\right)  \right]  \,,\;\mathbf{e}=(1,1,1)\,,\;U^{+}=U^{-1}\,.
\]

So far a few cases have been known when the two-level system admits exact
solutions, see the pioneer work of Rabi \cite{Rab37} where a spatially
homogeneous and time-dependent external magnetic field is analyzed, and the
work \cite{BagBaGW01} where exact solutions of the two-level system were found
for the following specific forms of potentials $f$:%
\begin{align}
f\left(  t\right)   &  =\frac{r_{0}}{\cosh\tau}\,,\;\tau=\frac{t}%
{T}\,,\label{1.4a}\\
f\left(  t\right)   &  =\frac{r_{0}}{T}\tanh\tau+\frac{r_{1}}{T}\,,
\label{1.4b}%
\end{align}
where $r_{0},\,r_{1}$ and $T$ are some constants. In what follows, we call the
potentials $f$ that admit exact solutions for the two-level system equations,
exact solvable potentials.

Sometimes, there exists the possibility to construct new exact solutions of
differential equations (in particular, of eigenvalue problems) with the help
of the Darboux transformation method \cite{Dar1456,MatSa91}. The idea of the
Darboux transformation method is to find an operator (an intertwining
operator) that relates solutions which correspond to different potentials.
Thus, if one knows solutions for a given potential, and a Darboux
transformation can be found, there exists a possibility to construct solutions
for another potential and at the same time to find this potential. The method
was applied for the first time by Darboux to find solutions of the
Sturm-Liouville problem. Applications of the Darboux transformations to
Schr\"{o}dinger-type equations can be found in the survey \cite{BagSa97}. For
the generalization of the method to sets of differential equations see e.g.
\cite{NiePeS03}.

In the present article, we adapt the Darboux procedure to the case of the
two-level system (Sect.II). In this respect, one ought to say that the Darboux
transformation for the general Zakharov-Shabat equations where studied in
\cite{MatSa91}, see also \cite{RogSc02}. However, such transformations cannot
be directly used in the case under consideration since, in the general case,
they violate the structure of the two-level system equations. We demonstrate
that one can construct the Darboux intertwining operator that does not violate
the specific structure of the equations of the two-level system, only
transforming potentials $f\left(  t\right)  $\ given by real functions into
other potentials of the same type. Then (Sect.III) we apply the obtained
Darboux transformation to known exact solutions of the two-level system. Thus,
we find three classes of new solutions for the two-level system and the
corresponding new potentials that allow such solutions.

\section{Darboux transformation for two-level system}

In this section, we adopt the Darboux procedure to equations (\ref{1},
\ref{5}). These equations can be also written as an eigenvalue problem as
follows:%
\begin{equation}
\hat{h}\Psi_{\varepsilon}=\varepsilon\Psi_{\varepsilon}\,,\;\hat{h}%
=i\sigma_{1}\frac{d}{dt}+V\left(  t\right)  \,,\;V\left(  t\right)
=i\sigma_{2}f\left(  t\right)  \,. \label{7}%
\end{equation}
Suppose we know solutions of the problem for any (complex) $\varepsilon$. And
suppose we can construct an intertwining operator $\hat{L}$, such that%
\begin{align}
&  \hat{L}\hat{h}=\hat{h}_{1}\hat{L}\,,\label{8}\\
&  \hat{h}_{1}=i\sigma_{1}\frac{d}{dt}+V_{1}\left(  t\right)  \,,\;V_{1}%
\left(  t\right)  =i\sigma_{2}f_{1}\left(  t\right)  \,. \label{9}%
\end{align}
Then the eigenvalue problem for the operator $\hat{h}_{1}$, has the following
solutions
\begin{equation}
\hat{h}_{1}\Phi_{\varepsilon}=\varepsilon\Phi_{\varepsilon}\,,\;\Phi
_{\varepsilon}=\hat{L}\Psi_{\varepsilon}\,. \label{10}%
\end{equation}
If the intertwining operator $\hat{L}$ is chosen to be%
\begin{equation}
\hat{L}=A\frac{d}{dt}+B\,, \label{12}%
\end{equation}
where $A\left(  t\right)  $ and $B\left(  t\right)  $ are some time dependent
$n\times n$ matrices, then the transformation from $\Psi_{\varepsilon}$ to
$\Phi_{\varepsilon}$ is called the Darboux transformation.

There is a general method of constructing the intertwining operators $\hat{L}$
(see for example \cite{NiePeS03} and references there) for a given eigenvalue
problem (\ref{7}). However, for our purposes the direct application of the
general method could not be useful. The point is that by application of this
method one can violate the specific structure of the initial potential $V_{0}%
$, that is the final potential $V_{1}$ will not have the specific structure
(\ref{8}) with a real function $f_{1}\left(  t\right)  $. Then, the final set
of equations is not a set of two-level system equations.

Thus, the peculiarity of our problem is that the matrix potentials $V_{0}$ and
$V_{1}$ must obey some algebraic restrictions and the Darboux transformation
has to respect these restrictions. In other words, we are looking for the
Darboux transformations that do not change the form of the equations of the
two-level system. The existence of such transformations is a nontrivial fact
which we are going to verify below.

The intertwining relation (\ref{7}) with the operator $\hat{L}$ in the form
(\ref{12}) and the potential $V_{1}$ in the form (\ref{9}) (such a choice is
always possible \cite{NiePeS03}) leads to the following relations%
\begin{align}
&  \sigma_{1}B-B\sigma_{1}+\sigma_{2}\left(  f_{1}-f\right)  =0\,,\label{13}\\
&  \sigma_{1}\dot{B}+\sigma_{2}Bf_{1}-\sigma_{2}\dot{f}-B\sigma_{2}f=0\,.
\label{14}%
\end{align}
Let us choose%
\begin{equation}
B=\alpha+i\left(  f-\beta\right)  \sigma_{3}\,, \label{15}%
\end{equation}
where $\alpha\left(  t\right)  $ and $\beta\left(  t\right)  $ are some real
functions. Then, we obtain for the function $f_{1}$,%
\begin{equation}
f_{1}=2\beta-f\,, \label{16}%
\end{equation}
and the equations for the functions $\alpha$ and $\beta$,%
\begin{equation}
\dot{\alpha}-2\beta\left(  f-\beta\right)  =0\,,\;\dot{\beta}+2\alpha\left(
f-\beta\right)  =0\,. \label{17}%
\end{equation}
One can easily see that there is a first integral of the equation (\ref{17})%
\begin{equation}
\alpha^{2}+\beta^{2}=R^{2}\,, \label{18}%
\end{equation}
where $R$ is a real constant. Note that (\ref{18}) is satisfied if we choose%
\begin{equation}
\alpha=R\cos\mu\,,\ \beta=f+R\sin\mu\,, \label{19}%
\end{equation}
with $\mu\left(  t\right)  $ a real function . Substituting (\ref{19}) into
(\ref{17}), we obtain for the function $\mu$ a transcendental differential
equation%
\begin{equation}
\dot{\mu}=2\left(  R\sin\mu-f\right)  \,. \label{20}%
\end{equation}
In what follows, we are going to find the functions $\alpha$ and $\beta$
independently, without the need to solve the equation (\ref{20}). Thus, at the
same time, we find in an indirect way solutions for the latter equation.

The time derivative in (\ref{12}) can be taken from equation (\ref{7}). Then
we obtain, with account taken of (\ref{15}),%
\begin{equation}
\Phi_{\varepsilon}=\left[  \alpha-i\left(  \varepsilon\sigma_{1}+\beta
\sigma_{3}\right)  \right]  \Psi_{\varepsilon}\,. \label{21}%
\end{equation}
Thus, we see that the Darboux transformation (for equations (\ref{1},
\ref{5})) that respects the restriction (\ref{9}) does exist. It has the
algebraic form (\ref{21}) and is determined by solutions of equations
(\ref{17}), or by equations (\ref{20}). To finish the construction, one has to
be able to represent solutions of the set (\ref{17}) with the help of
solutions of the initial equations (\ref{1}, \ref{5}). Such a possibility
exists and is described below.

Let us introduce the spinor $\bar{\Psi}$,%
\begin{equation}
\bar{\Psi}=-i\sigma_{2}\Psi^{\ast}=\left(
\begin{array}
[c]{c}%
-\psi_{2}^{\ast}\\
\psi_{1}^{\ast}%
\end{array}
\right)  \,,\;\left(  \bar{\Psi},\Psi\right)  =0\,, \label{23}%
\end{equation}
where $\ast$ denotes complex conjugation. As follows from (\ref{1}), this
spinor obeys the equation%
\begin{equation}
i\overset{\cdot}{\bar{\Psi}}=\left(  \mathbf{\sigma F}^{\ast}\right)
\bar{\Psi}\,,\;\mathbf{F}^{\ast}=\left(  \varepsilon^{\ast},0,f\right)  \,.
\label{24}%
\end{equation}
In addition, we introduce a complex vector $\mathbf{p}\left(  t\right)  $,%
\begin{equation}
\mathbf{p}=\left(  \bar{\Psi},\mathbf{\sigma}\Psi\right)  \,. \label{25}%
\end{equation}
With account taken of (\ref{23}), we can see that for any spinor $\Psi$, the
relation%
\begin{equation}
\mathbf{p}^{2}=p_{1}^{2}+p_{2}^{2}+p_{3}^{2}=\left(  \bar{\Psi},\Psi\right)
^{2}=0 \label{26}%
\end{equation}
holds true. An equation for the vector $\mathbf{p}$ follows from (\ref{1}) and
(\ref{24}),%
\begin{equation}
\mathbf{\dot{p}}=2\left[  \mathbf{F}\times\mathbf{p}\right]  \,. \label{28}%
\end{equation}
Let us introduce the functions $\alpha$ and $\beta$ as%
\begin{equation}
\alpha=-\varepsilon\frac{p_{2}}{p_{3}}\,,\;\beta=-\varepsilon\frac{p_{1}%
}{p_{3}}\,. \label{29}%
\end{equation}
In virtue of (\ref{26}), these functions are related as
\begin{equation}
\alpha^{2}+\beta^{2}=-\varepsilon^{2}\,. \label{30}%
\end{equation}
It is easy to verify (using (\ref{28}) and (\ref{30})) that the functions
$\alpha$ and $\beta$ obey the set of equations (\ref{17}). Setting
$\varepsilon=-iR$ in (\ref{23})-(\ref{30}), we can conclude that (\ref{30})
coincides with (\ref{18}). In addition, it follows from (\ref{28}) that there
exist such solutions of this equation such that $p_{1}$ and $p_{2}$ are real,
whereas $p_{3}$ is imaginary and can be determined from (\ref{30}). This
provides the reality of the functions $\alpha$ and $\beta$.

Thus, we have expressed solutions of the set (\ref{17}) via solutions of the
initial equations (\ref{5}) at $\varepsilon=-iR$, where $R$ is real.
Substituting (\ref{29}) into (\ref{21}), one can find the final form of both
the Darboux transformation for equations (\ref{1}, \ref{5}) and the function
$f_{1}\left(  t\right)  $ (which determines the new potential, see
(\ref{9})),
\begin{align}
&  \Phi_{\varepsilon}=q^{-1}\sigma_{2}\left(  \mathbf{\sigma\tilde{p}}\right)
\Psi_{\varepsilon}\,,\;\mathbf{\tilde{p}}=\left(  p_{1},p_{2},\varepsilon
q/R\right)  \,,\nonumber\\
&  f_{1}=-2\frac{p_{1}}{q}-f\,,\;q=q\left(  t\right)  =\sqrt{p_{1}^{2}%
+p_{2}^{2}}\,. \label{31}%
\end{align}
Here the components $p_{1}$ and $p_{2}$ are constructed with the help of the
equations (\ref{25}) via solutions of equations (\ref{1}, \ref{5}) at
$\varepsilon=-iR$ with a real $R$. We stress that the constructed
transformation preserves the form invariance of equations (\ref{1}, \ref{5}).
All the components of the transformation can be constructed in an algebraic
manner via solutions of the initial equation.

\section{New exact solutions for two-level systems}

In the following, we apply the above consideration to obtain new solutions of
equations (\ref{1}, \ref{5}).

\subsection{The first case}

Let $f\left(  t\right)  =c_{0}=$ $\mathrm{const}$ in equations (\ref{1},
\ref{5}). The corresponding solutions have the form%

\begin{align}
\psi_{1}  &  =i\left(  c_{0}-\omega\right)  p_{0}\exp\left(  i\omega t\right)
-\varepsilon q_{0}\exp\left(  -i\omega t\right)  \,,\nonumber\\
\psi_{2}  &  =i\varepsilon p_{0}\exp\left(  i\omega t\right)  +\left(
c_{0}-\omega\right)  q_{0}\exp\left(  -i\omega t\right)  \,,\nonumber\\
\omega^{2}  &  =\sqrt{c_{0}^{2}+\varepsilon^{2}}\,, \label{32}%
\end{align}
where $q_{0}$ and $p_{0}$ are arbitrary complex constants. Using these
solutions at $\varepsilon=-iR_{0}$ we find (using the above procedure) the
functions $\alpha=\alpha_{0}\left(  t\right)  $ and $\beta=\beta_{0}\left(
t\right)  $ as well as the corresponding potential $f_{1}\left(  t\right)  $:%
\begin{align}
&  \alpha_{0}\left(  t\right)  =-\frac{\dot{Q}_{0}}{2\left(  Q_{0}%
+c_{0}\right)  }\,,\;\beta_{0}\left(  t\right)  =c_{0}+\frac{R_{0}^{2}%
-c_{0}^{2}}{Q_{0}+c_{0}}\,,\nonumber\\
&  Q_{0}=Q_{0}\left(  t\right)  =\left\{
\begin{array}
[c]{c}%
R_{0}\cosh\varphi_{0}\,,\;R_{0}^{2}>c_{0}^{2}\,\\
R_{0}\cos\varphi_{0}\,,\;R_{0}^{2}<c_{0}^{2}\,
\end{array}
\right.  \,,\nonumber\\
&  \varphi_{0}=2\left(  \omega_{0}t+\gamma_{0}\right)  \,,\;\omega_{0}%
=\sqrt{\left|  R_{0}^{2}-c_{0}^{2}\right|  }\,. \label{33}%
\end{align}
Here $\gamma_{0}$ is an arbitrary real constant.

The corresponding potential $f_{1}\left(  t\right)  $ reads:%
\begin{equation}
f_{1}\left(  t\right)  =2\beta_{0}\left(  t\right)  -c_{0}=c_{0}%
+\frac{2\left(  R_{0}^{2}-c_{0}^{2}\right)  }{Q_{0}+c_{0}}\,. \label{34}%
\end{equation}
For $c_{0}\neq0$ the function $f_{1}\left(  t\right)  $ is not a particular
case of the potential (\ref{1.4a}) and is a new solvable potential for
equations (\ref{1}, \ref{5}).

The spinor $\Phi_{\varepsilon}$ can be easily constructed according to formula
(\ref{21}). We do not exhibit here its explicit form which is too cumbersome.

\subsection{The second case}

The Darboux transformation can be applied again to the exact solution obtained
in the previous section. We represent here only the functions $\alpha
_{1}\left(  t\right)  $ and $\beta_{1}\left(  t\right)  $ of such a
transformation with $\varepsilon=-iR_{1}$. They are:%

\begin{align}
\alpha_{1}\left(  t\right)  =  &  R_{1}S\left[  2\alpha_{0}\left(  c_{0}%
\beta_{0}-R_{1}^{2}\right)  Q_{1}+\left(  2\beta_{0}^{2}-R_{0}^{2}-R_{1}%
^{2}\right)  \dot{Q}_{1}/2+2R_{1}\alpha_{0}\left(  \beta_{0}-c_{0}\right)
\right]  \,,\nonumber\\
\beta_{1}\left(  t\right)  =  &  -R_{1}S\left\{  \left[  c_{0}\left(
2\beta_{0}^{2}-R_{0}^{2}+R_{1}^{2}\right)  -2\beta_{0}R_{1}^{2}\right]
Q_{1}-\alpha_{0}\beta_{0}\dot{Q}_{1}\right. \nonumber\\
&  \left.  +R_{1}\left[  R_{1}^{2}-R_{0}^{2}+2\beta_{0}\left(  \beta_{0}%
-c_{0}\right)  \right]  \right\}  \,, \label{35}%
\end{align}
where%
\begin{align*}
&  S^{-1}=R_{1}\left(  R_{0}^{2}+R_{1}^{2}-2\beta_{0}c_{0}\right)
Q_{1}+\alpha_{0}R_{1}\dot{Q}_{1}+\left(  R_{0}^{2}+R_{1}^{2}\right)
c_{0}-2\beta_{0}R_{1}^{2}\,,\\
&  Q_{1}=\left\{
\begin{array}
[c]{c}%
R_{1}\cosh\varphi_{1}\,,\;R_{1}^{2}>c_{0}^{2}\,,\\
R_{1}\cos\varphi_{1}\,,\;R_{1}^{2}<c_{0}^{2}\,,
\end{array}
\right. \\
&  \varphi_{1}=2\left(  \omega_{1}t+\gamma_{1}\right)  \,,\;\omega_{1}%
=\sqrt{\left|  R_{1}^{2}-c_{0}^{2}\right|  }\,.
\end{align*}
Here the functions $\alpha_{0}\left(  t\right)  $ and $\beta_{0}\left(
t\right)  $ are defined by equations (\ref{33}) and $\gamma_{1}$ is a real constant.

The corresponding solvable potential $f_{1}^{\prime}\left(  t\right)  $ has
the form:%
\begin{equation}
f_{1}^{\prime}\left(  t\right)  =2\beta_{1}\left(  t\right)  +f_{1}\left(
t\right)  =2\left[  \beta_{1}\left(  t\right)  +\beta_{0}\left(  t\right)
\right]  -c_{0}\,. \label{36}%
\end{equation}
We stress that for nonzero $R_{0}$ and $R_{1}$ this potential is not a
particular case of any known exact solvable potentials.

As before, the spinor $\Phi_{\varepsilon}$ can be easily constructed according
to formula (\ref{21}) and is not exhibited here.

\subsection{The third case}

Now we assume that the function $f$ has the form (\ref{1.4b}). In this case,
solutions of (\ref{1}, \ref{5}) can be written as (see \cite{BagBaGW01})%
\begin{align}
\psi_{1}=  &  \left(  1-z\right)  ^{\nu}E\left[  c_{1}z^{\mu}F\left(
a+1,b;c;z\right)  +c_{2}z^{-\mu}F\left(  \bar{a}+1,\bar{b};\bar{c};z\right)
\right]  \,,\nonumber\\
\psi_{2}=  &  \left(  1-z\right)  ^{\nu}\left[  \left(  r_{0}-r_{1}%
+2i\mu\right)  c_{1}z^{\mu}F\left(  a,b+1;c;z\right)  +\right. \nonumber\\
&  \left.  \left(  r_{0}-r_{1}-2i\mu_{0}\right)  c_{2}z^{-\mu}F\left(  \bar
{a},\bar{b}+1;\bar{c};z\right)  \right]  \,, \label{3.9}%
\end{align}
where%
\begin{align*}
&  z=\frac{1}{2}\left(  1+\tanh\tau\right)  \,,\;a=\mu+\nu+ir_{0}%
\,,\;b=\mu+\nu-ir_{0}\,,\;\bar{a}=-\mu+\nu+ir_{0}\,,\\
&  \bar{b}=-\mu+\nu-ir_{0}\,,\;\,c=1+2\mu,\;\bar{c}=1-2\mu,\;E=\varepsilon
T\,,
\end{align*}
and $c_{1}$ and $c_{2}$ are complex constants. If the following relations are
satisfied%
\begin{align}
4\mu^{2}+E^{2}+\left(  r_{0}-r_{1}\right)  ^{2}  &  =0\,,\nonumber\\
4\nu^{2}+E^{2}+\left(  r_{0}+r_{1}\right)  ^{2}  &  =0\,, \label{3.10}%
\end{align}
we can identify $F\left(  a,b;c;z\right)  $ with the hyper-geometrical function.

We are going to construct the operator $\hat{L}$ in the case where $\mu$ and
$\nu$ are real. Therefore, setting $E=-iR$ in (\ref{3.10}), the reality
condition will be satisfied if%
\[
R^{2}>\max\left(  r_{0}\pm r_{1}\right)  ^{2}\,.
\]
In this case, we can write%
\begin{equation}
\mu_{0}=\frac{1}{2}\sqrt{R^{2}-\left(  r_{0}-r_{1}\right)  ^{2}}\,,\;\;\nu
_{0}=\frac{1}{2}\sqrt{R^{2}-\left(  r_{0}+r_{1}\right)  ^{2}}\,, \label{3.11}%
\end{equation}
and the expressions (\ref{3.9}) become%
\begin{align}
&  \psi_{1}^{\left(  0\right)  }=-iR\left(  1-z\right)  ^{\nu_{0}}\left(
c_{1}z^{\mu_{0}}F_{0}+c_{2}z^{-\mu_{0}}F_{1}\right)  \,,\nonumber\\
&  \psi_{2}^{\left(  0\right)  }=\left(  1-z\right)  ^{\nu_{0}}\left[  \left(
r_{0}-r_{1}+2i\mu_{0}\right)  c_{1}z^{\mu_{0}}F_{0}^{\ast}+\left(  r_{0}%
-r_{1}-2i\mu_{0}\right)  c_{2}z^{-\mu_{0}}F_{1}^{\ast}\right]  \,,\nonumber\\
&  F_{0}=F\left(  a_{0}+1,a_{0}^{\ast};1+2\mu_{0};z\right)  \;,\quad
F_{1}=F\left(  \bar{a}_{0}+1,\bar{a}_{0}^{\ast};1-2\mu_{0};z\right)
\,,\nonumber\\
&  a_{0}=\mu_{0}+\nu_{0}+ir_{0}\;,\quad\bar{a}=-\mu_{0}+\nu_{0}+ir_{0}\,.
\label{3.12}%
\end{align}
The constants $c_{1}$ and $c_{2}$ will be chosen such that the relation%
\begin{equation}
\frac{c_{1}}{c_{2}}=p^{2\mu_{0}}\left(  r_{0}-r_{1}-2i\mu_{0}\right)
R^{-1}=p^{2\mu_{0}}e^{-2i\varphi_{0}} \label{3.13}%
\end{equation}
is satisfied, where $p$ is a new real constant and $\varphi_{0}$ is a constant
phase defined, in agreement with (\ref{3.11}), by the expression%
\begin{equation}
\left(  r_{0}-r_{1}+2i\mu_{0}\right)  R^{-1}=e^{2i\varphi_{0}}\,. \label{3.14}%
\end{equation}
For such a choice of the constants $c_{1}$ and $c_{2},$ the solutions
(\ref{3.12}) assume the form%
\begin{align}
&  \psi_{1}^{\left(  0\right)  }=-iR\left(  1-z\right)  ^{\nu_{0}}\sqrt
{c_{1}c_{2}}A\,,\;\psi_{2}^{\left(  0\right)  }=R\left(  1-z\right)  ^{\nu
_{0}}\sqrt{c_{1}c_{2}}A^{\ast}\,,\nonumber\\
&  A=\left(  pz\right)  ^{\mu_{0}}e^{-i\varphi_{0}}F_{0}+\left(  pz\right)
^{-\mu_{0}}e^{i\varphi_{0}}F_{1}\,. \label{3.15}%
\end{align}
Using the above solutions (\ref{3.12}) and (\ref{3.15}) and the expression
(\ref{25}) and (\ref{29}), the constants $\alpha$ and $\beta$ are seen to be
real, and they can be written as%
\begin{equation}
\alpha=\frac{iR\left(  A^{\ast2}-A^{2}\right)  }{2TAA^{\ast}}\,,\;\beta
=f+\frac{R\left(  A^{\ast2}+A^{2}\right)  }{2TAA^{\ast}}\,, \label{3.16}%
\end{equation}
with $f$ defined by (\ref{1.4b}). In this case, the Darboux transformation
provides exact solutions of equations (\ref{1}, \ref{5}) with a potential
given by%
\begin{equation}
f_{1}=\frac{R\left(  A^{\ast2}+A^{2}\right)  }{T\left|  A\right|  ^{2}}%
-\frac{r_{0}}{T}\tanh\tau-\frac{r_{1}}{T}\,. \label{3.18}%
\end{equation}
We remark that the new potential, as well as the corresponding solutions, are
only expressed via the hyper-geometric functions.

\begin{acknowledgement}
V.G.B thanks FAPESP for support and Nuclear Physics Department of S\~{a}o
Paulo University for hospitality, as well as he thanks Russia President grant
SS-1743.2003.2 and RFBR grant 03-02-17615 for partial support; M.C.B. thanks
FAPESP and D.M.G thanks both FAPESP and CNPq for permanent support.
\end{acknowledgement}

\end{document}